\documentclass[%
  onecolumn,
  oneside,
  amsmath,amssymb,
  aps,
  pra,
  nofootinbib,
  superscriptaddress,
  a4paper
]{revtex4-2}

\usepackage{graphicx}%
\usepackage[usenames,dvipsnames]{xcolor}
\usepackage{siunitx}
\usepackage{physics}  
\usepackage{natbib}

\usepackage[
centering, includefoot,
text={6.0in,10.2in},
total={6.3in,8.75in},
top=0.8in, left=1.15in,
]{geometry}

\usepackage[
  bookmarks=false,
  colorlinks,
  linkcolor=blue,
  urlcolor=blue,
  citecolor=blue,
  plainpages=false,
  pdfpagelabels,
  final,
  breaklinks=true
]{hyperref}
\hypersetup{
pdftitle={Title}, 
pdfauthor={Emilio Pisanty}
}

\DeclareSIUnit{\au}{{a.u.}}

\def\fig#1{Fig.\,\ref{#1}}

\def\beq#1{\begin{equation}\label{#1}}
\def\eeq{\end{equation}}

\begin{document}

\title{Macroscopic effects in generation of attosecond XUV pulses \\ via high-order frequency mixing in gases and plasma}

\author{V.~A.~Birulia}
\affiliation{Moscow Institute of Physics and Technology (National Research University), Dolgoprudny, Moscow Region 141700, Russia}

\author{M.~A.~Khokhlova}
\affiliation{King’s College London, Strand Campus, London WC2R 2LS, UK}
\email{margarita.khokhlova@kcl.ac.uk}

\author{V.~V.~Strelkov}
\affiliation{Prokhorov General Physics Institute of the Russian Academy of Sciences, Vavilova street 38, Moscow 119991, Russia}
\affiliation{Moscow Institute of Physics and Technology (National Research University), Dolgoprudny, Moscow Region 141700, Russia}

\begin{abstract}
We study the generation of attosecond XUV pulses via high-order frequency mixing~(HFM) of two intense generating fields, and compare this process with the more common high-order harmonic generation~(HHG) process. We calculate the macroscopic XUV signal by numerically integrating the 1D propagation equation coupled with the 3D time-dependent Schr\"odinger equation. We analy\-tically find the length scales which limit the quadratic growth of the HFM macroscopic signal with propagation length. Compared to HHG these length scales are much longer for a group of HFM components, with orders defined by the frequencies of the generating fields. This results in a higher HFM macroscopic signal despite the microscopic response being lower than for HHG. In our numerical simulations, the intensity of the HFM signal is several times higher than that for HHG in a gas, and it is up to three orders of magnitude higher for generation in plasma; it is also higher for longer generating pulses. The HFM provides very narrow XUV lines ($\delta \omega / \omega = 4.6 \times 10^{-4}$) with well-defined frequencies, thus allowing for a simple extension of optical frequency standards to the XUV range. Finally, we show that the group of HFM components effectively generated due to macroscopic effects provides a train of attosecond pulses such that the carrier-envelope phase of an individual attosecond pulse can be easily controlled by tuning the phase of one of the generating fields.
\end{abstract}

\maketitle
\noindent

\section{Introduction}
The generation of attosecond XUV pulses is a task of historically fundamental and currently expanding applied and technological importance. This creates strong demand for the generation of brighter, shorter and more controllable attosecond pulses. 

Traditionally the table-top production of attosecond pulses is associated with the process of high-harmonic generation~(HHG)~\cite{Corkum2007, Krausz2009, Villeneuve2018, Ryabikin2023} in gaseous media, when a strong laser field interacts with atoms or molecules of the gas in a highly-nonlinear nonperturbative fashion, giving rise to the emission of a wide plateau of laser harmonics stretching into the XUV region, including the water window~\cite{Johnson2018Aug} and even keV photons~\cite{Popmintchev2012}. This XUV emission, in the temporal representation, takes the shape of a train of attosecond pulses generated every half-cycle of the laser field. The brightness of the attosecond pulses can be boosted by either optimizing the atomic response by carefully choosing the generating field and the gas type~\cite{Chipperfield2009, Raz2012}, or by taking advantage of the phase-matched generation from many atoms~\cite{Kim2008, Weissenbilder2022} (of course one can combine both).

It is known that HHG suffers from phase matching and there is a limit for the propagation distance, where the efficient coherent (quadratic) growth of the HHG signal with propagation changes to a slower one. Moreover, recently we have found~\cite{KhokhlovaStrelkov, KhokhlovaStrelkov2023}, analytically and numerically, a fundamental limit, the blue-shift length, where the energy of harmonics generated at high intensities of the laser switches its quadratic growth to a linear one. This limit appears due to the ionization of the medium resulting in the blue shift of the driver and, consequently, of the harmonics. The limited HHG efficiency hinders technological progress and the widening of the application field. One of the ways to overcome this limitation is switching to a different process which is fundamentally close to HHG but suffers less from the phase-matching problem~--- high-order frequency mixing~(HFM).

HFM occurs when a weak low-frequency field is added to the strong laser one, which has an intensity comparable to those used for HHG. As a result, there are new frequency components emitted: the generation of an XUV photon involves many photons from the laser field and a few photons from the low-frequency field. Within the single-particle response, next to each high harmonic in the emitted spectrum there are a few satellites that appear on both sides. In the macroscopic response only red-shifted satellites survive the phase matching~\cite{Hort2021}, and these perform a quadratic growth far beyond the high harmonics. 

This favorable coherent growth of the HFM signal with propagation length is, first, rooted in the behavior of its microscopic response. The behavior of the phases and intensities of the microscopic response as functions of the parameters of the driving fields have been studied theoretically and experimentally~\cite{Eichmann1995, Gaarde1996, Balcou1999} and the prospect of phase-matched HFM was shown already in the early studies~\cite{Shkolnikov, Milchberg, Chichkov_plasma, Shkolnikov_1996, Platonenko_non-collinear, Chichkov_2000}. In particular, it was demonstrated~\cite{Shkolnikov,Milchberg,Chichkov_plasma} that using drivers with a certain frequency ratio allows the phase-matched generation of specific harmonics of relatively low order in plasma, which was followed by further studies in this direction~\cite{Kapteyn_2007,Worner_non-collinear, PhysRevLett.112.143902, Oguchi_PRA, Strelkov_np, Ellis_2017, Tran_2019, Harkema:19, Ganeev_2016, KhokhlovaStrelkov}. 

The photoionization of the generating medium inevitably accompanies rescattering processes, so even for the generation in gases the plasma dispersion is important and can be dominating. In this case for HFM by two fields with strong-field frequency $\omega_0$ and weak-field frequency  
\beq{freq1}
\omega_1=\frac{|m|}{Q} \omega_0 \, ,
\eeq
where $Q$ is a natural number , $m$ is an integer number, and $Q+m$ is odd, the dephasing due to plasma dispersion is zero for the generation of the HFM component with the frequency
\beq{HFM_freq}
\omega_{Q,m}= Q \omega_0 + m \omega_1 
\eeq
for $m<0$. Better phase matching leads to a high macroscopic response for HFM~\cite{KhokhlovaStrelkov}, while the HFM microscopic response is lower than the one for HHG. The photoionization of the medium provides the frequency shift of the driving fields, the so-called plasma-induced blue shift. Assuming that the shift is small in comparison with the frequencies, there is no frequency shift for the generation of the HFM $(Q,m)$ component~\cite{Hort2021}, and the frequency shift is also small for HFM components $(q,m)$ for integer orders $q$ that neighbor the ideally-phase matched $Q$. 

The scope of this paper is the detailed study of the macroscopic properties of the HFM response, which take into account the temporal variation of the phase matching and the frequency shifts of the generating fields. Here we find the ratio of the HFM intensity to the HHG intensity for the generation in a gas and in plasma with initial ionization degree equal to unity. 

We also show that one can expect very narrow XUV lines emitted in the HFM process, which originate from the phase-matched generation under varying plasma density (i.e.\ during a relatively long time), and the compensation of the blue shift. The investigation of this prospect motivates another direction of our studies.

Another promising property we discovered in the microscopic response for HFM is that the carrier-envelope phase~(CEP) of attopulses generated via HFM changes from pulse to pulse in the train in a way that is defined by the frequencies of the strong and weak fields and the order of the satellites. Moreover, this CEP can be easily controlled by the phases of the driving fields~\cite{Birulia2022}. Here we study whether this control is feasible for the macroscopic response.

We perform a detailed study of the macroscopic properties of the attosecond pulses generated via HFM. We find out that a group of HFM components with several orders $q \approx Q$ is generated satisfying the phase-matching conditions, thereby this natural selection of the specific group of HFM components (the order of harmonic satellites) with high generation efficiency (due to the phase matching) leads to the production of a bright attosecond pulse train. 

We study the macroscopic response of HFM by numerically integrating the propagation equation for the fields in a 1D geometry. While the nonlinear polarization of the medium is calculated by numerically solving the 3D time-dependent Schr\"odinger equation~(TDSE) for a single-active electron~(SAE) atom in the field at each propagation step.

\section{Theory}
\subsection{Spatial scales for the HHG macroscopic response}
The well-known spatial scales describing the HHG macroscopic response are the XUV absorption length $L^{(q)}_\mathrm{abs}$, where $q$~is the harmonic order, and the coherence length $L^{(q)}_\mathrm{coh}$~\cite{Constant}. The latter is inversely proportional to the phase mismatch between the generating and generated fields: $ L_\mathrm{coh}^{(q)}={\pi}/{|\Delta k_{q}|}$, where $\Delta {k}_q=q k_0-k_q$ is the phase mismatch for the generation $q$-th harmonic, $k_0$ and $k_q$ are the wavevectors of the fundamental and harmonic fields, respectively. Below we assume that the medium dispersion makes the dominant contribution to the phase mismatch and that the refractive index for the harmonic field is unity. Then the coherence length is defined by the difference of the refractive index for the fundamental from unity, $\Delta n$, and it is written as
\beq{L_coh}
L_\mathrm{coh}^{(q)}=\frac{\pi c}{q\omega_0 \Delta n} = \frac{ \lambda_q}{2 \Delta n } \, .
\eeq

However, in the HHG process under high laser intensities the phase mismatch varies in time due to the medium ionization. If this variation is comparable to the mismatch itself, the coherence length essentially varies within the duration of the generation. In this case the macroscopic properties of the generation of $q$-th harmonic can be better characterized with the blue-shift length~\cite{KhokhlovaStrelkov, KhokhlovaStrelkov2023}
\beq{L_bs}
L_\mathrm{bs}^{(q)}= \frac{\pi c}{q \omega_0 |n_\mathrm{f}-n_\mathrm{i}|} = \frac{ \lambda_q}{2 |n_\mathrm{f}-n_\mathrm{i}| } \, ,
\eeq
where $n_\mathrm{i}$ and $n_\mathrm{f}$ are the initial and final values of the refractive index of the fundamental field. 

This result is obtained considering phase matching via direct integration of the generated fields over both temporal and spatial coordinates. This theory assumes 1D~propagation and linear change in refractive index with time, and it neglects the modification of the fundamental-pulse envelope during the propagation in the medium~\cite{KhokhlovaStrelkov}. There is also a simple explanation: at the propagation distances up to $L_\mathrm{bs}^{(q)}$ the harmonic generation is phase-matched, so the harmonic signal grows quadratically with propagation length. For longer propagation distances the ionization-induced blue shift of the fundamental field $\delta \omega_0$ leads to the blue shift of the local single-atom harmonic response $\delta \omega_q = q \delta \omega_0$, which exceeds the initial harmonic bandwidth ($\approx \pi/ \tau$, where $\tau$ is the pulse duration of a single harmonic). As a result, the fields generated at different propagation lengths do not add coherently any more, the XUV intensity stops growing, and the bandwidth of the macroscopic HHG response increases linearly, providing linear growth of the total XUV energy. For the studies in the next sections we note that the harmonic blue shift after propagation over $L_\mathrm{bs}^{(q)}$ is equal to the harmonic bandwidth
\beq{harmonic_bs}
\delta \omega_q(L_\mathrm{bs}^{(q)}) \approx \frac{\pi}{\tau} \, ,
\eeq
where
\beq{harmonic_vs_fundamental_bs}
\delta \omega_q (L) =  q \, \delta \omega_0 (L).
\eeq

\subsection{Coherence length for HFM}
Let us consider the spatial scales for the HFM process, starting from the coherence length. For the HFM $(q,m)$ component the coherence length is
\beq{L_coh_q,m_def}
L_\mathrm{coh}^{(q,m)}=\frac{\pi}{|\Delta {k}^{(q,m)}|} \, ,
\eeq
where $\Delta k^{(q, m)}=q k_0+m k_1-k^{(q,m)}$ is the phase mismatch for the generation of the $(q,m)$-th component, ${k}_0$, ${k}_1$ and $k^{(q, m)}$ are the wavevectors of the initial fields and the generated HFM field, respectively. If the plasma and/or capillary contribution to the dispersion dominates and the frequencies of the fields $\omega_{0,1}$ sufficiently exceed the plasma frequency $\omega_\mathrm{pl}$, then the refractive index for frequency $\omega$ is $n=1-\omega_\mathrm{pl}^2/(2 \omega^2)$ and the detuning from the phase matching for the process is written as
\begin{equation}
\Delta k^{(q,m)}=\frac{\omega_\mathrm{pl}^2}{2c} \left( -\frac{q}{\omega_0}-\frac{m}{\omega_1}+\frac{1}{q \omega_0+m \omega_1}\right) \, .
\label{eq:Delta_k_}
\end{equation}

Neglecting the last term in~\eqref{eq:Delta_k_} for high $q$, one can see from Eq.~\eqref{eq:Delta_k_} and Eq.~\eqref{freq1} that for $q=Q$ the coherence length~\eqref{L_coh_q,m} is infinite, and for $q \ne Q$ we have
\beq{L_coh_q,m}
L_\mathrm{coh}^{(q,m)}=L_\mathrm{coh}^{(Q)} \frac{Q}{|\Delta q|} \, ,
\eeq
where $\Delta q = q - Q$.

\subsection{Blue-shift length for HFM}
We define the HFM blue-shift length $L_\mathrm{bs}^{(q,m)}$ following similar logic as for the HHG blue-shift length, described above: the frequency shift of the HFM microscopic response $\delta \omega_{q,m}(L_\mathrm{bs}^{(q,m)})$ is equal to its initial bandwidth
\beq{condition}
|\delta \omega_{q,m}(L_\mathrm{bs}^{(q,m)})|= \pi/ \tau \, .
\eeq
where, analogously to Eqs.~\eqref{harmonic_bs} and \eqref{harmonic_vs_fundamental_bs},
\beq{blue-shifted_HFM}
\delta \omega_{q,m}(L)= q \, \delta \omega_0 (L) + m \, \delta \omega_1 (L) \, .
\eeq

For further studies we focus on $\delta \omega_0 (L)$ and $\delta \omega_1 (L)$. Note that considering the blue shift of the fundamental, one can assume that it is much smaller than the fundamental frequency; however, this is not necessarily the case for the low-frequency field. Therefore, below we take into account the plasma-induced blue shift more carefully than it is usually done for strong laser fields.

Below we omit indexes $0$ and $1$ for the fundamental and low-frequency fields. So $\omega$ is the initial field frequency and $\tilde \omega (L)$ is the blue-shifted frequency of the field, then $\tilde \omega (L) \equiv \omega+ \delta \omega (L)$. 

When the field propagates over distance $dL$, its phase advance is
$$
\delta \varphi (t)= k(t) dL \, ,
$$
where $k=n \tilde \omega /c$ is a wavenumber. Then the frequency shift is
$$
\delta \omega \equiv \frac{\partial}{\partial t} \delta \varphi (t) = \frac{\partial k}{\partial t} dL \, .
$$
Taking into account that 
$\frac{\partial \tilde \omega }{\partial L}= \frac{\partial }{\partial L} \delta \omega$, we have
$$
\frac{\partial \tilde \omega }{\partial L} =\frac{1}{c} \frac{\partial }{\partial t} \left[ n \tilde \omega\right] \, .
$$
Substituting 
\beq{ref_ind}
n=1-\omega_\mathrm{pl}^2/(2 \tilde \omega^2) \, ,
\eeq
we get 
\beq{dif_eq_1}
\frac{\partial \tilde \omega }{\partial L} =\frac{1}{c} \left[ \frac{\partial \tilde \omega}{\partial t} \left( 1+ \frac{\omega_\mathrm{pl}^2}{2 \tilde \omega^2 }\right)  
 - \frac{1}{2 \tilde \omega} \frac{\partial \omega_\mathrm{pl}^2}{\partial t}   \right] \, .
\eeq

The squared plasma frequency is proportional to the density of free electrons $N_e$, or $\omega_\mathrm{pl}^2 \propto N_e$. Assuming that this density varies {\it linearly} with time (which is very reasonable near the peak of the pulse), we find that the factor $\partial \omega_\mathrm{pl}^2 / \partial t$ in the last term in Eq.~\eqref{dif_eq_1} does not depend on time. In this case, this equation reduces to
$$
\frac{\partial \tilde \omega }{\partial t}=0 \, ,
$$
\beq{dif_eq_2}
\frac{\partial \tilde \omega }{\partial L} = - \frac{1}{2 \tilde \omega c} \frac{\partial \omega_\mathrm{pl}^2}{\partial t} \, .
\eeq
Introducing the constant $\alpha$ as
\beq{alpha}
\frac{\partial \omega_\mathrm{pl}^2}{\partial t}=- \alpha c \, ,
\eeq
we rewrite Eq.~\eqref{dif_eq_2} as
\beq{dif_eq_3}
\frac{\partial \tilde \omega }{\partial L} =  \frac{\alpha }{2 \tilde \omega} \, ,
\eeq
where the solution is
\beq{omega_L}
\tilde \omega (L)=  \sqrt{\omega^2+ \alpha L} \, .
\eeq

Assuming that, for both the fundamental and the low-frequency fields, the frequency variation is small in comparison with the frequency itself
\beq{condition1}
\delta \omega = \tilde \omega -\omega \ll \omega \, ,
\eeq
we expand the square root in Eq.~\eqref{omega_L} as 
\beq{root}
\sqrt{\omega^2+ \alpha L} \approx \omega + \frac{\alpha L }{2 \omega} \, ,
\eeq
then we have
\beq{lin}
\delta \omega_\mathrm{lin} = \frac{\alpha L }{2 \omega} \, ,
\eeq
where index $\mathrm{lin}$ denotes the linear approximation~\eqref{root}.

Now we go back to restoring the indices of the fundamental field,~$0$, and of the low-frequency one,~$1$. From Eq.~\eqref{lin} we can write the constant $\alpha$ for the fundamental field as
\beq{0_lin}
\delta \omega_{0,\mathrm{lin}} = \frac{\alpha L }{2 \omega_0} \, .
\eeq
Combining this equation with Eqs.~\eqref{harmonic_bs} and \eqref{harmonic_vs_fundamental_bs}, we find
\beq{alpha_vs_L_bs}
\alpha = \frac{2 \pi \omega_0 }{ \tau Q L_\mathrm{bs}^{(Q)}} \, ,    
\eeq
here we would like to point out that $\alpha$ does not depend on $Q$ because $Q L_\mathrm{bs}^{(Q)}$ does not depend on $Q$.

In this way, we can find the blue-shift length for the $(q,m)$ component in the linear approximation. For that, we substitute $\alpha$~\eqref{alpha_vs_L_bs}, which is the same for the fundamental and for the weak field, into Eq.~\eqref{lin}. Then we insert $\delta \omega_{0,\mathrm{lin}}$ and $\delta \omega_{1,\mathrm{lin}}$~\eqref{lin} into Eqs.~\eqref{condition} and \eqref{blue-shifted_HFM} to obtain
\beq{L_bs_q,m_lin}
L_\mathrm{bs, lin}^{(q,m)}=L_\mathrm{bs}^{(Q)} \frac{Q}{|\Delta q|} \, .
\eeq
Note that this equation is similar to Eq.~\eqref{L_coh_q,m}, and they both predict infinite lengths of the quadratic growth for $q=Q$. 

To consider $q$ close to $Q$ more accurately, one should consider the blue shift of the low-frequency field beyond the linear approximation~\eqref{root}. Here it is worth noting that for the fundamental field this approximation is very reliable, so we continue using it. However, for the low-frequency field, its frequency is much closer to the plasma frequency than the fundamental; therefore, the low-frequency field reacts much more strongly to changes in the plasma frequency. Thus, using Eq.~\eqref{condition1} for the low-frequency field with Eqs.~\eqref{alpha_vs_L_bs} and \eqref{freq1}, we get
\beq{delta_omega_1}
\delta \omega_1(L)=\sqrt{\omega_1^2+\alpha L}-\omega_1=\omega_1\left(\sqrt{1+\frac{2 \pi L}{|m|\tau \omega_1 L_\mathrm{bs}^{(Q)}} }-1\right) \, .
\eeq
Then substituting $\delta \omega_{0,\mathrm{lin}}(L)$~\eqref{0_lin} and $\delta \omega_1(L)$~\eqref{delta_omega_1} into Eqs.~\eqref{blue-shifted_HFM} and \eqref{condition}, we get the following equation for negative $m$ values
\beq{abs_eq}
\abs{\frac{q}{Q}\frac{L_\mathrm{bs}^{(q,m)}}{L_\mathrm{bs}^{(Q)}}-2{\cal N}^2\left(\sqrt{1+\frac{1}{{\cal N}^2}\frac{L_\mathrm{bs}^{(q,m)}}{L_\mathrm{bs}^{(Q)}}}-1\right)}=1 \, ,
\eeq
where
\beq{N}
 {\cal N}=\sqrt{\frac{ |m| \tau}{T_1}} 
\eeq
and $T_1=2\pi/\omega_1$ is an optical cycle of the low-frequency field. 

Note that for positive $m$ `plus' sign must replace `minus' one in the left-hand side of Eq.~\eqref{abs_eq} and appear before parenthesis. Moreover, for $|m|=1$ the parameter ${\cal N}$ means the square root of the number of cycles of the low-frequency field within time $\tau$ (which is close to the duration of the fundamental pulse). 

From Eq.~\eqref{abs_eq} we can find the ratio ${L_\mathrm{bs}^{(q,m)}}/{L_\mathrm{bs}^{(Q)}}$ as
\beq{L_bs_q,m_full}
\frac{L_\mathrm{bs}^{(q,m)}}{L_\mathrm{bs}^{(Q)}}=
\begin{cases}
\frac{Q}{q}\left(1-2{\cal N}^2\frac{\Delta q}{q}\pm 2{\cal N} \sqrt{\left({\cal N}\frac{\Delta q}{q}\right)^2 +\frac{Q}{q}} \right) \, ,
\\
\\
\frac{Q}{q}\left(-1-2{\cal N}^2\frac{\Delta q}{q}\pm 2{\cal N} \sqrt{\left({\cal N}\frac{\Delta q}{q}\right)^2-\frac{Q}{q}} \right) \, .
\end{cases}
\eeq

For $\Delta q /q \ll 1$ and ${\cal N}\Delta q /q \ll 1$ (practically ${\cal N}$ is of the order of unity, so these conditions are similar to each other) we can simplify the upper solution in Eq.~\eqref{L_bs_q,m_full} with the `plus' sign in front of the square root. Using the expression $Q=q-\Delta q$ and expanding the root in a linear Taylor series over $\Delta q /q$, we obtain
\beq{L_bs_q,m_small_case}
\frac{L_\mathrm{bs}^{(q,m)}}{L_\mathrm{bs}^{(Q)}}=\left(1-\frac{\Delta q}{q}\right)\left(1-2{\cal N}^2\frac{\Delta q}{q} + 2{\cal N} \left[1+\frac{1}{2}\left({\cal N}\frac{\Delta q}{q}\right)^2 -\frac{1}{2}\frac{\Delta q}{q}
+\ldots \right] \right) \, ,
\eeq
and then by neglecting terms proportional to $(\Delta q /q)^2$ and their greater powers, we get
\beq{L_bs_q,m}
L_\mathrm{bs}^{(q,m)}\approx L_\mathrm{bs}^{(Q)} \left[1 +2 {\cal N} - \frac{\Delta q}{q} \left(1+3{\cal N}+ 2 {\cal N}^2\right)\right] \, .
\eeq
We do not consider the alternative upper solution, which incorporates `minus' prior to the square root, as it results in a negative outcome for $L_\mathrm{bs}^{(q,m)}$ for small $\Delta q$ values. Similarly, in the lower solutions~\eqref{L_bs_q,m_full}, we encounter a problematic scenario where a negative value appears under the root again for small~$\Delta q$.

In the opposite case of high $\Delta q$ (or when ${\cal N}\Delta q /q \gg 1$), we can simplify Eq.~\eqref{L_bs_q,m_full} to Eq.~\eqref{L_bs_q,m_lin}. To do this, we rewrite Eq.~\eqref{L_bs_q,m_full} as follows:
\beq{L_bs_q,m_full_for_big_delta_q}
\frac{L_\mathrm{bs}^{(q,m)}}{L_\mathrm{bs}^{(Q)}}=
\begin{cases}
\frac{Q}{q}\left(1-2{\cal N}^2\frac{\Delta q}{q}\pm 2{\cal N}^2\frac{|\Delta q|}{q} \sqrt{1+\frac{1}{{\cal N}^2}\frac{qQ}{\Delta q^2}} \right) \, ,
\\
\\
\frac{Q}{q}\left(-1-2{\cal N}^2\frac{\Delta q}{q}\pm 2{\cal N}^2\frac{|\Delta q|}{q} \sqrt{1-\frac{1}{{\cal N}^2}\frac{qQ}{\Delta q^2}} \right) \, .
\end{cases}
\eeq
Expanding the root in the Taylor series and leaving the first two terms, we get the desired expression~\eqref{L_bs_q,m_lin}. From several solutions for $L_\mathrm{bs}^{(q,m)}$, one should choose the minimum positive value as the physical solution.

\subsection{HHG and HFM macroscopic generation efficiency}
In the HHG and HFM processes, the photoionization of the medium makes a significant contribution to the refractive index, which leads to the non-stationary phase matching. If the ionization is strong enough, then this renders $L_\mathrm{coh}$ ineffective as a length scale descriptor for phase matching due to its variation during the laser pulse. 

Additionally, the photoionization induces a blue shift in the laser frequency. Albeit small, when multiplied by the harmonic order, this shift noticeably impacts the high-harmonic line. When the blue shift exceeds the initial harmonic bandwidth, the generated XUV no longer combines coherently with the initial XUV, eliminating both constructive and destructive interference, thereby stopping the quadratic intensity growth. Thus, in this case, $L_\mathrm{bs}$ is more suitable for the description of the effective length. 

One can conclude that the spatial scale of the quadratic growth for the HFM field is defined by the shortest of the lengths given by Eqs.~\eqref{L_coh_q,m}, \eqref{L_bs_q,m_lin} and \eqref{L_bs_q,m}. As a result, the intensity of the HFM components is proportional to the square of the propagation length until it reaches the first limit. From these equations, we see that HFM can be more efficient than HHG for $q \approx Q$ (for example, the coherence length and the blue-shift length in the linear approximation for HFM are $|Q/\Delta q|$ times greater than for the harmonic $q=Q$ in HHG). For this group of most effectively generated HFM components the limitation is given by Eq.~\eqref{L_bs_q,m}.

From this equation we see that for negative $\Delta q$ (i.e.\ $q<Q$) the blue-shift length is longer than for the positive ones. This can be understood as following: the saturation of the root dependence~\eqref{delta_omega_1} with propagation leads to slower growth of $\delta \omega_{1}$ for long propagation distances, and for $q<Q$ this is compensated by lower term $q \delta \omega_{0}$ in the total shift~\eqref{blue-shifted_HFM}. Thus, several HFM components below $Q$ should be generated with highest efficiency. Moreover, from Eq.~\eqref{L_bs_q,m} one can see that $L_\mathrm{bs}^{(q,m)}$ is longer (thus the HFM efficiency is higher) for long fundamental pulses, which still satisfy the condition ${\cal N}\Delta q /q \ll 1$. Therefore, in the calculations below we are using relatively long pulses.

The efficiency of HFM depends also on the microscopic response rate. By choosing a favorable intensity of the low-frequency field one can maximize this rate. This maximal intensity of the HFM microscopic response can be roughly estimated as follows. In the HFM scenario, the HHG microscopic response intensity of the $q$-th harmonic is redistributed among several HFM $(q,m)$ components. By optimizing the intensity of the low-frequency field, it is possible to achieve the case when $2|m|+1$ HFM components generate with roughly equal intensities. 

For instance, at $|m|=1$, we observe the generation of the odd $q$-th harmonic and two peaks near the even harmonic $(q-1, \pm1)$, and at $|m|=2$, in addition to the former, two peaks near the odd harmonic $(q,\pm2)$ are produced, and so on. Consequently, the intensity of the optimized HFM microscopic response decreases with $|m|$ as $1/(2|m|+1)$. 

On the other hand, as derived from Eq.~\eqref{L_bs_q,m}, $(L_\mathrm{bs}^{(q,m)})^2$ is proportional to ${\cal N}^2=|m| \tau /T_1$~\eqref{N} for the case when $q=Q$. Taking into account $T_1= 2 \pi / \omega_1$ and Eq.~\eqref{freq1}, we finally find $\left( L_\mathrm{bs}^{(q,m)} \right)^2 \propto m^2$. This trade-off between microscopic response intensity and the square of the blue-shift length indicates that higher $|m|$ is more preferable. However, very long propagation lengths are impractical both computationally and experimentally due to heavy calculations and absorption, respectively. For that reason, in this paper we focus on the HFM process with $|m|=2$.

\section{Numerical simulation of the propagation}
We numerically simulate the propagation of the laser field in argon gas through the direct integration of the reduced wave equation for the external field in a 1D~geometry. The nonlinear polarization, which enters the wave equation, is also calculated numerically by solving the 3D TDSE for an atom (in the SAE approximation) in the field at each propagation step. The method we use is described in detail in~\cite{KhokhlovaStrelkov, KhokhlovaStrelkov2023}.

In particular, to calculate the polarization of a gas target we calculate the expectation value of the time-dependent total force  $f_z(x, t)$~\footnote{In Ref.~\cite{KhokhlovaStrelkov}, there is a misprint in this equation.} as 
\begin{equation}
f_z(x,t)=E(x,t)-\expval{\nabla V(\textbf{r})}{\Psi(\textbf{r},t)} \, ,   
\label{gas_polarization}
\end{equation}
where $\Psi(\textbf{r},t)$ is the wavefunction found via the numerical TDSE solution at the progressing propagation step. In contrast to~\cite{KhokhlovaStrelkov, KhokhlovaStrelkov2023} here we consider also the propagation in plasma. Assuming the generation in plasma plumes (see Sec.~\ref{plasma}), we consider a medium which is preionized by a heating laser pulse, where the preionization degree is unity. The main (driving) laser field further ionizes it. The force (per one ion) consists of a force acting at the bound electron and a force acting at the electron, which became free due to the preionization. The former force is given by Eq.~\eqref{gas_polarization} and the latter one is just $E(x,t)$. So we have 
\begin{equation}
f_z^\mathrm{(pl)}(x,t)=2 E(x,t)-\expval{\nabla V(\textbf{r})}{\Psi(\textbf{r},t)} \, .    
\label{plasma_polarization}
\end{equation}

Finally, we should note that the results as functions of the propagation distance are presented for a certain choice of medium density. The outcomes actually depend on the product of the length and the medium density, so they can be obviously rescaled for another density. However, our approach for the simulation of the propagation assumes that the electronic density is less than the critical plasma density for all considered propagating fields, or, in other words, that the frequency of the low-frequency field is higher than the plasma frequency 
\begin{equation}
    \omega_1 > \omega_\mathrm{pl} \, .
    \label{pl}
\end{equation}

\section{Results}
\subsection{HFM in a gas}
\begin{figure}
\centering
\includegraphics[width=0.65\linewidth]{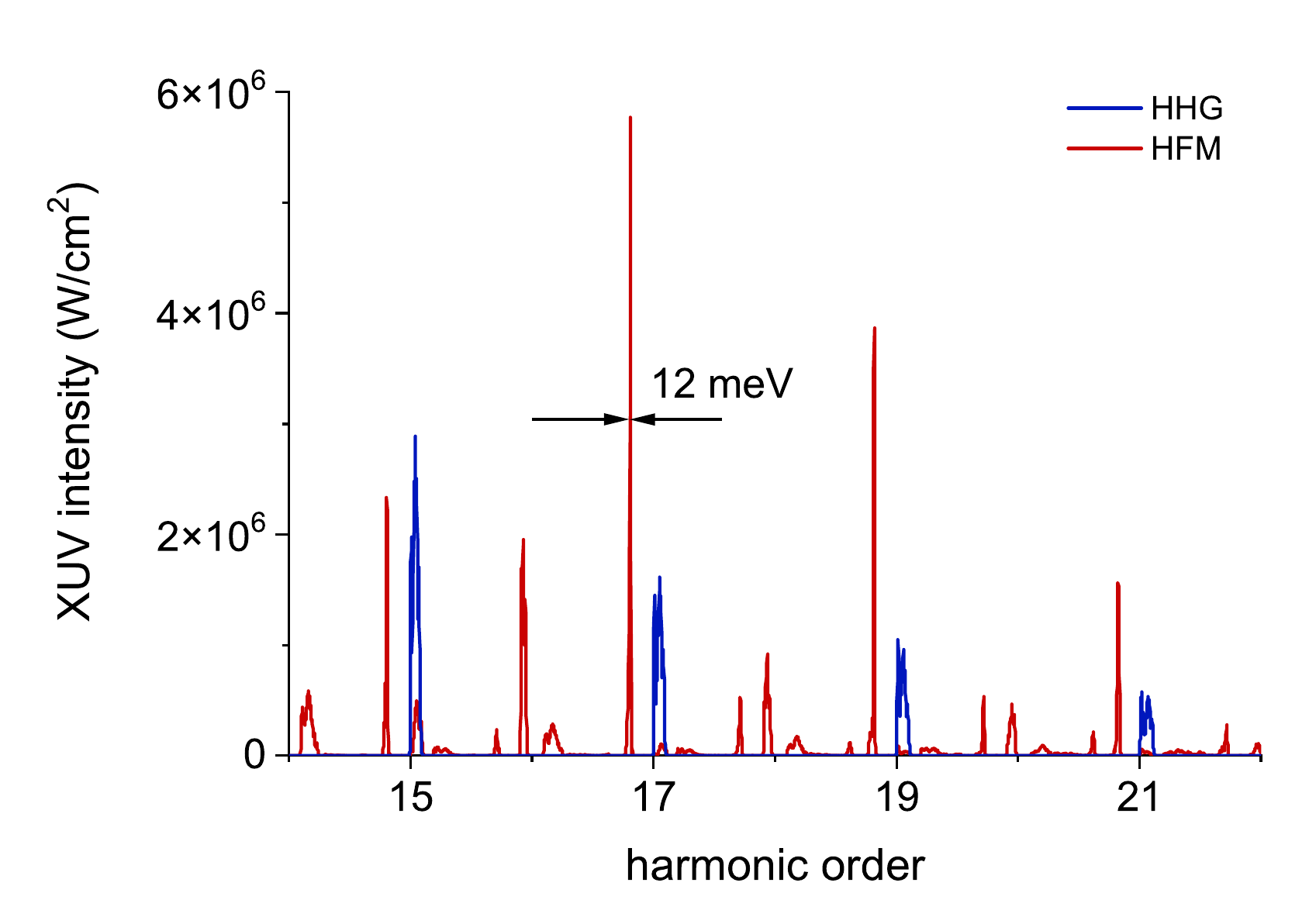}
\caption{Spectra of the macroscopic response for the HHG~(blue) and HFM~(red) processes. The fundamental intensity is $\SI{1.6e14}{W/cm^2}$, the wavelength is \SI{800}{nm} and the pulse duration~(FWHM) is \SI{500}{fs}. The weak low-frequency field intensity is $2\%$ from the fundamental one, the wavelength is $\SI{8.4}{\micro m}$ ($\omega_1=\frac{2}{21}\omega_0$) and the pulse duration is the same as for the fundamental field. The density of the argon gas is \SI{3.0e18}{cm^{-3}}. The propagation length is \SI{1.2}{mm}. The simulation for HHG is done for the case when only the fundamental field propagates in the medium.}
\label{Spectrum}
\end{figure}

Here we present numerical results for HFM in argon gas. The spectra emitted in the case of HHG and HFM are shown in~\fig{Spectrum}. The calculation parameters are presented in the caption; these parameters are used throughout this section if not specified otherwise. From~\fig{Spectrum} one can see that for the chosen initial parameters of the incident fields the HFM components with $m=-2$ are effectively generated for several neighboring $q$. Moreover, a very narrow XUV line (FWHM is \SI{12}{meV}) is generated due to the HFM process with $q=17$ and $m=-2$. This linewidth is close to the inverse duration of the microscopic response in the pulse with duration \SI{500}{fs} (hence, broader than this inverse duration). The possibilities of using such narrow line as an XUV frequency standard are described in~Sec.~\ref{Discussion}.

\begin{figure}
\centering
\includegraphics[width=0.65\linewidth]{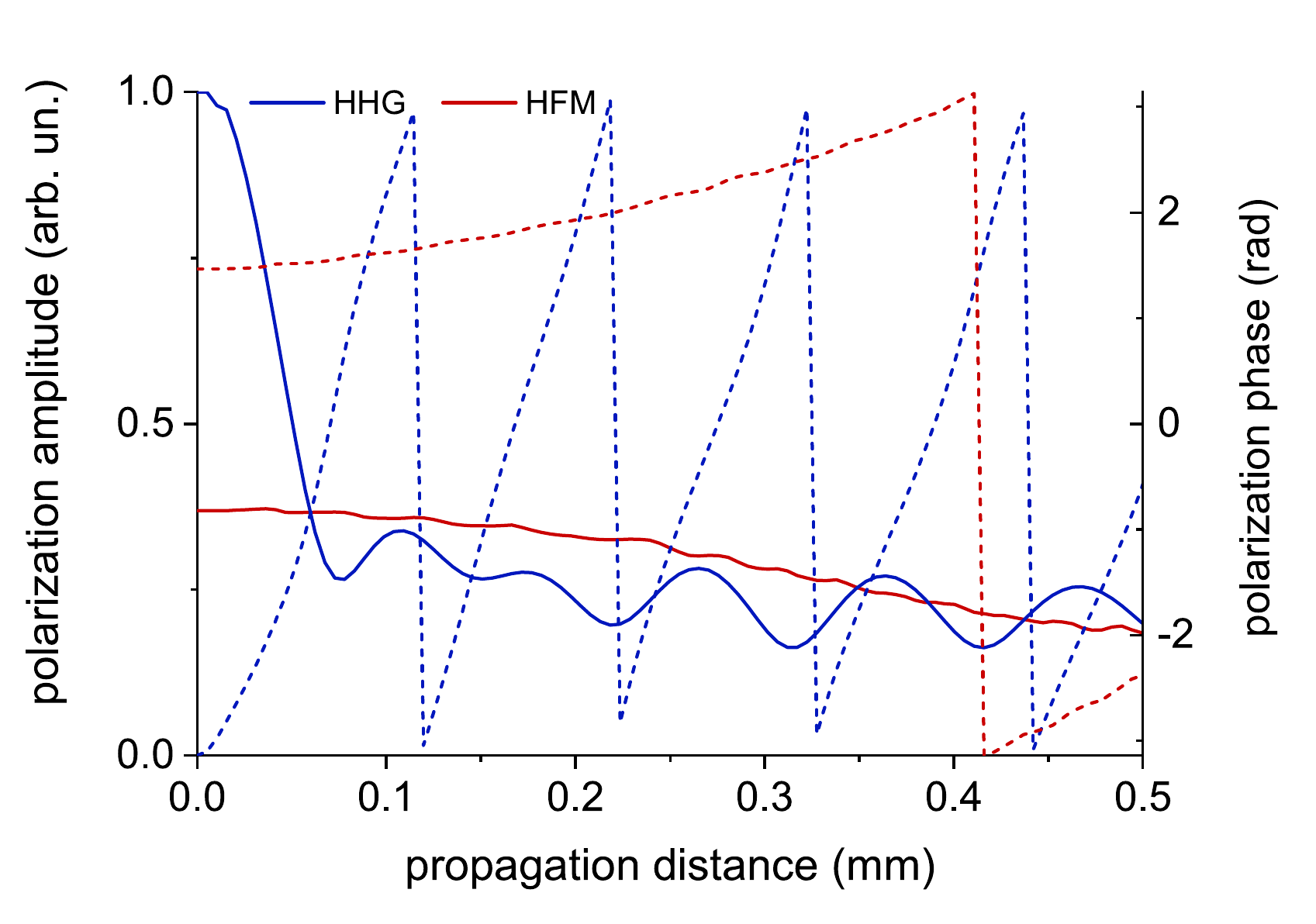}
\caption{Amplitudes~(solid curves) and phases~(dashed curves) of the microscopic response as functions of the propagation length for H21~(blue) and HFM component with $q=21$, $m=-2$~(red).}
\label{Microscopic}
\end{figure}

To study the differences in the HHG and HFM responses in more detail, we first compare the microscopic responses as functions of the propagation length in~\fig{Microscopic}. Here we calculate the properties of the microscopic response for the central frequencies of the HHG and HFM components. In~\fig{Microscopic} we see that initially the microscopic response for the HFM process is lower. We optimize the low-frequency field intensity to maximize the $q=21, m=-2$ response. In these conditions the H21 `response intensity' is approximately equally distributed among five HFM components: $q=21, m=0$; $q=20, m=\pm 1$; $q=21, m=\pm 2$. Thus, the intensity of the microscopic response of the HFM component with $q=21, m=-2$ is approximately $1/5$ of the intensity of the HHG microscopic response.

\fig{Microscopic} also shows that both the HHG and HFM amplitudes decrease with propagation length. This decrease mainly occurs due to the blue shifts of the generating fields. For HHG this comes from the deep ionization of the medium (specifically, at 40\%) by the field, so the blue-shift length limits the growth of the signal. For the HFM component shown in the figure we have $\Delta q=0$, thus the coherence length given by Eq.~\eqref{L_coh_q,m} is infinite and the generation is also limited by the blue shift. In this figure we see that the HFM microscopic response does not vanish up to the longer propagation length than the HHG one. Namely, the HHG amplitude falls down to half of its maximum after approximately \SI{0.05}{mm} of propagation, and the HFM amplitude falls down to half of its maximum after \SI{0.5}{mm} of propagation. The ratio of these lengths is close to the analytical prediction~\eqref{L_bs_q,m} giving $L_\mathrm{bs}^{(q,m)}/L_\mathrm{bs}^{(Q)} =12.8$ for these conditions ($q=Q=21$, $|m|=2$, $\tau = \SI{500}{fs}$). Moreover, we see that the phase of the microscopic response is stable for HFM and this is not the case for HHG. 

\begin{figure}
\centering
\includegraphics[width=0.65\linewidth]{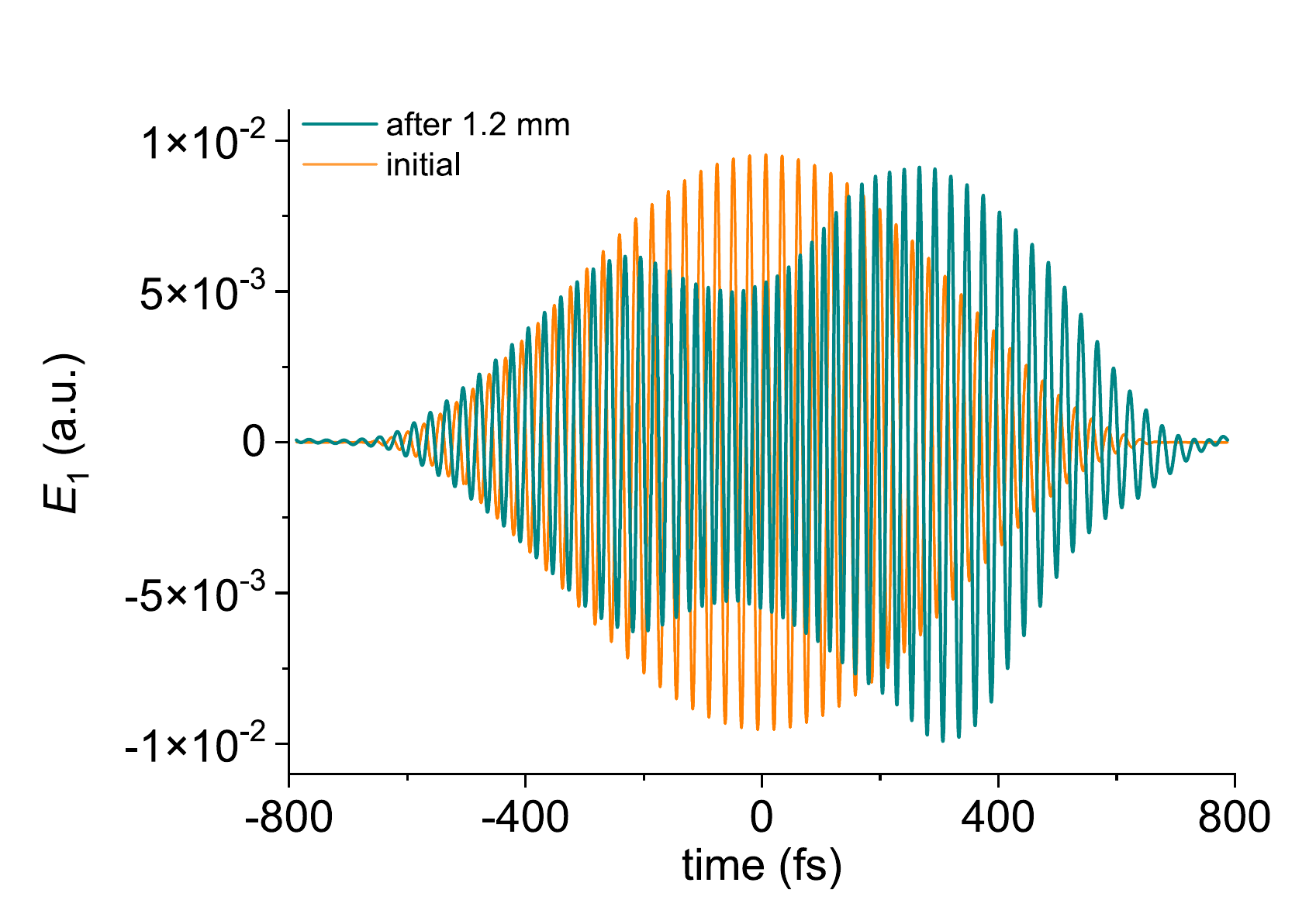}
\caption{Modification of the low-frequency generating pulse with propagation distance.}
\label{low_freq_pulse}
\end{figure}

One more reason for the decrease of the HFM response with propagation length is the nonlinear modification of the low-frequency generating field, see~\fig{low_freq_pulse}. Namely, the intensity of the low-frequency field becomes lower in the central part of the pulse, where the XUV emission is generated, and the HFM response decreases with the descending intensity of the low-frequency field.  

\begin{figure}
\centering
\includegraphics[width=0.65\linewidth]{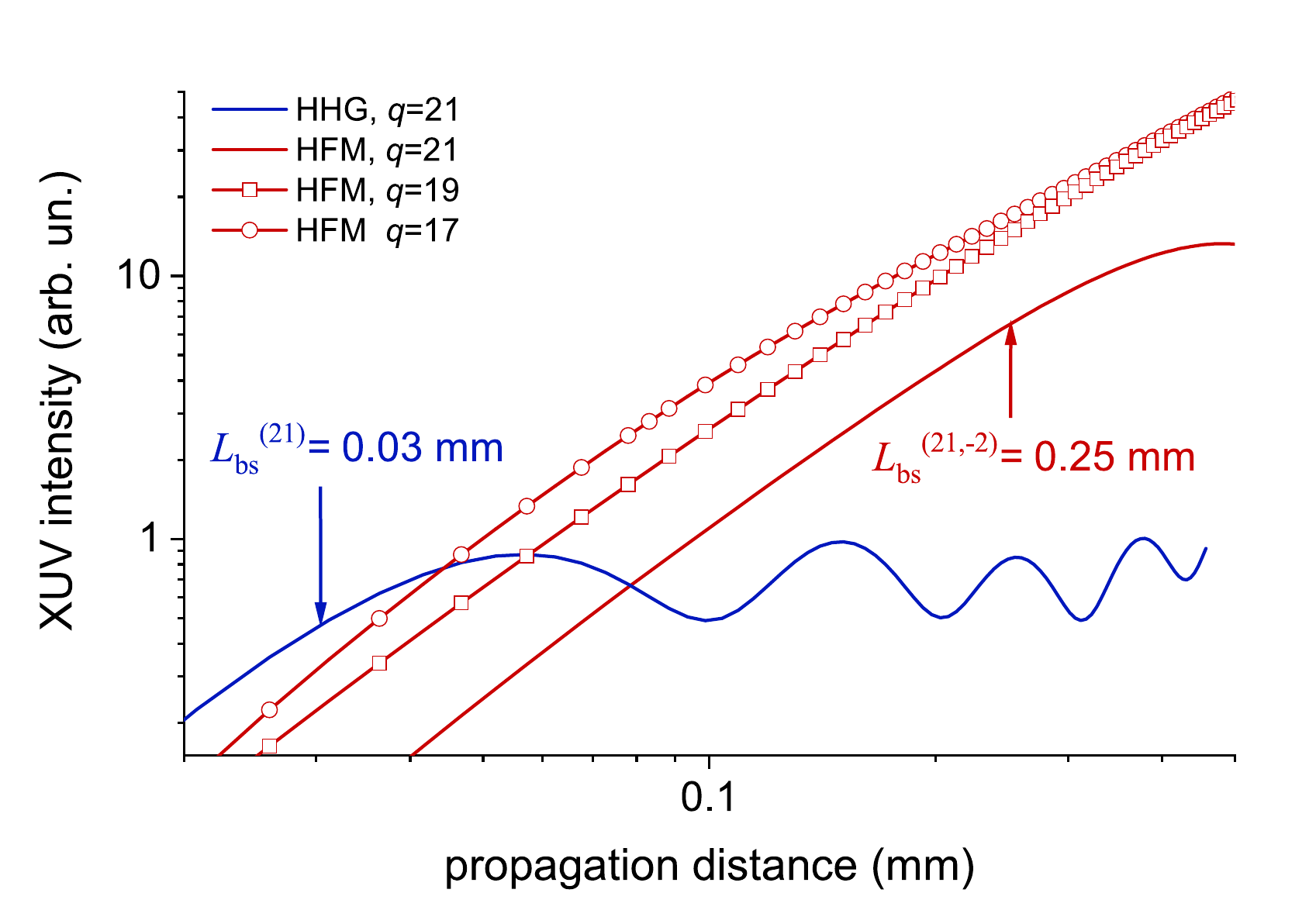}
\caption{Intensity of the macroscopic response as a function of the propagation length for H21~(blue), and for HFM components with $q=17$, $19$, $21$ and $m=-2$~(red).}
\label{Macroscopic}
\end{figure}

Now we focus on the comparison of the macroscopic responses. \fig{Macroscopic} shows the intensities of the macroscopic signal for HHG and HFM with the same $m=-2$ and several $q$. One can see that for short propagation distances the HHG signal is much more intense, due to the higher microscopic polarization amplitude, see Fig.~\ref{Microscopic}. Initially the intensity of the HHG signal grows quadratically and then saturates, in agreement with findings of~\cite{KhokhlovaStrelkov2023}. The HFM signals continue growing and saturate at much longer distances due to both phase mismatch and blue shift of the fields. Thus, for long propagation distances the HFM signals are more intense than the HHG one.

\begin{figure}
\centering
\includegraphics[width=0.65\linewidth]{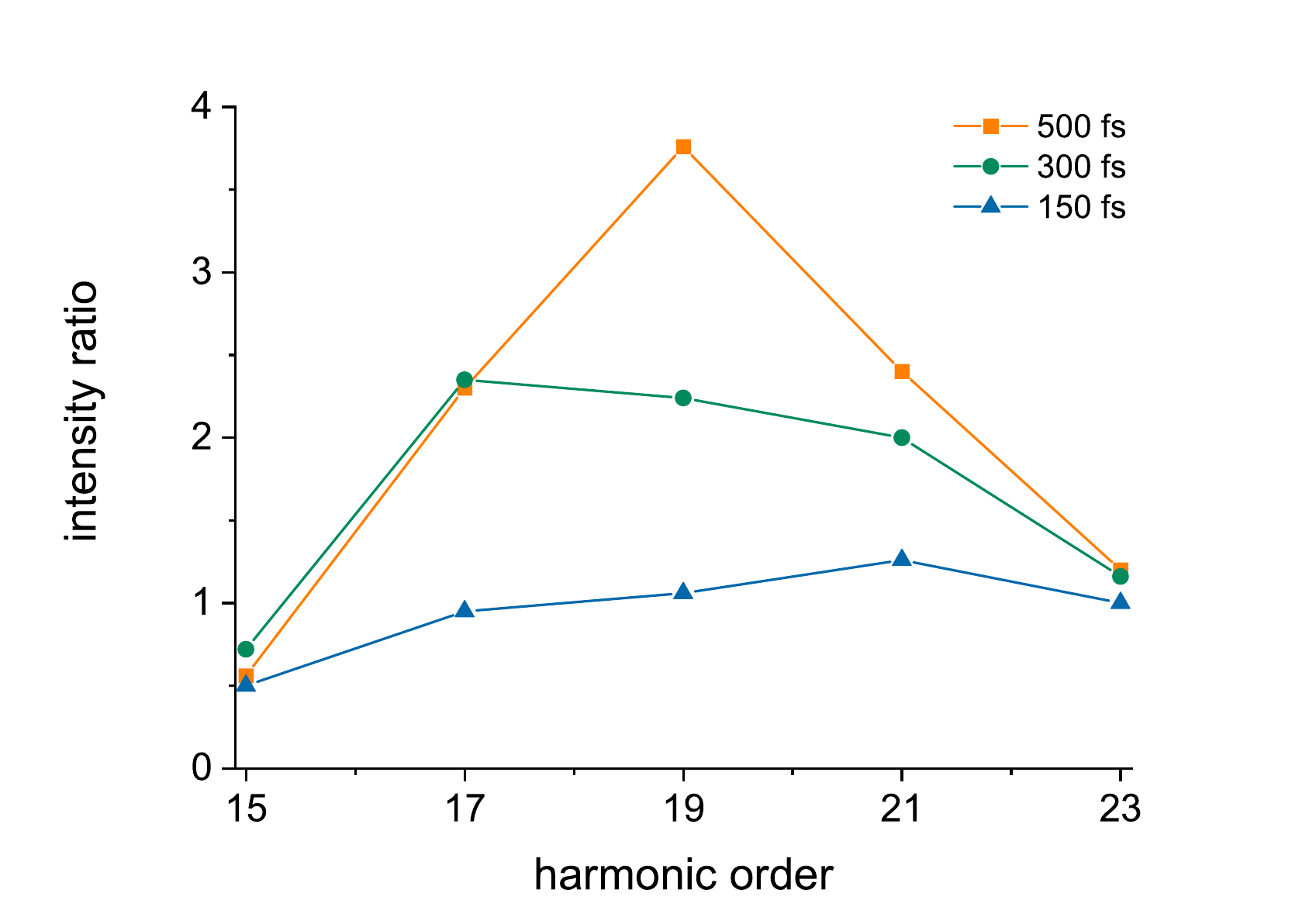}
\caption{Ratio of the peak intensities between the HFM ($m=-2$) and HHG signals after \SI{1.2}{mm} of propagation. The calculation is done for several pulse durations shown in the graph. The laser intensity is $\SI{1.5e14}{W/cm^2}$, the low-frequency field intensity is 2\% of this value, other parameters are the same as above.}
\label{Gain_gas}
\end{figure}

We also calculate the ratio of the peak intensities between the HFM signal and the HHG signal. \fig{Gain_gas} shows the ratio of the peak intensity of the HFM component to the peak intensity of the high harmonic with the same $q$, thus it shows `the gain' achieved in HFM in comparison to HHG. 

\subsection{HFM in plasma}\label{plasma}
Here we compare the HHG and HFM efficiencies in plasma and study prospects of the attosecond pulse production via such processes. HHG in this scenario is actively studied experimentally using plasma plumes at metallic surfaces as the generating medium~\cite{Ganeev2006, Singh2023, Mathijssen2023} (for a review see~\cite{Ganeev2013}). Using plasma for HHG allows, in particular, the resonant enhancement of the microscopic response~\cite{Ganeev2006, Singh2021, Armstrong2021}. However, the free-electron dispersion leads to a very short coherence length. This problem can be solved, to some extent, by quasi-phase matching~\cite{Ganeev2014, Strelkov2017}. Another way is the realization of phase-matched HFM using two generating fields with frequencies given by Eq.~\eqref{freq1}. 

For phase matching in gases the condition~\eqref{pl} is easily satisfied, however, for a plasma medium it provides an essential limitation on the plasma density and the frequency of the weak field $\omega_1$. The plasma density, which we use in this section, is lower than the gas density we consider for the gas HFM calculations. Namely, the initial ionization degree of plasma is unity and the plasma density is \SI{1.0e18}{cm^{-3}}; note that usually it is about or below this value~\cite{Strelkov2017} in experiments with laser plumes. For such density and $\omega_1=2/21 \omega_0=2/21 \omega_\mathrm{Ti:Sapp}$ the requirement~\eqref{pl} is fulfilled.

In common plasma HHG experiments the initial ionization degree in the plasma plume is unity, and the degree of further ionization by the laser pulse is much smaller, thus we can write that
\beq{plasma_index}
|n_\mathrm{f}-n_\mathrm{i}| \ll |1-n_\mathrm{i}| \, .
\eeq
In this case $\Delta n \approx |1-n_\mathrm{i}|$ and from Eqs.~\eqref{L_coh} and \eqref{L_bs} one can devise that $L_\mathrm{coh}^{(q)}<L_\mathrm{bs}^{(q)}$ for HHG for all orders $q$, or using the terminology of Refs.~\cite{KhokhlovaStrelkov, KhokhlovaStrelkov2023}, the HHG in plasma is phase-matching limited. From Eqs.~\eqref{L_coh_q,m} and \eqref{L_bs_q,m_lin} one can see that this occurs for all HFM components {\it except} the component  $(Q, -|m|)$, while the coherence length for this component is infinity. Thus, the quadratic growth of this very component is limited by the blue-shift length~\eqref{L_bs_q,m}, which is written, taking into account that $\Delta q =0$, as
\beq{L_bs_q,m_plasma}
L_\mathrm{bs}^{(Q,m)}\approx L_\mathrm{bs}^{(Q)} \left[1 +2 {\cal N} \right]= \frac{ \lambda_q}{2 |n_\mathrm{f}-n_\mathrm{i}| }\left[1 +2 {\cal N} \right] \, .
\eeq

Comparing Eqs.~\eqref{L_coh} and \eqref{L_bs_q,m_plasma}, and taking into account Eq.~\eqref{plasma_index}, one concludes that the length of quadratic growth of the generation efficiency for the HFM $(Q, -|m|)$ component is much longer than for high harmonics of any order. Note that this applies even to low $\cal{N}$, therefore below we use moderate pulse duration of \SI{50}{fs}. For the neighboring HFM components ($\Delta q \ne 0$, $|\Delta q | \ll Q$) the length of the quadratic growth is the coherent length~\eqref{L_coh_q,m}. It is shorter than the length of quadratic growth for $\Delta q =0$ (given by Eq.~\eqref{L_bs_q,m_plasma}), but longer than this length for HHG. Thus, for moderate propagation distances the group of HFM components is efficiently generated, while for longer ones a single HFM component ($\Delta q =0$) keeps growing.

\begin{figure}
\centering
\includegraphics[width=0.65\linewidth]{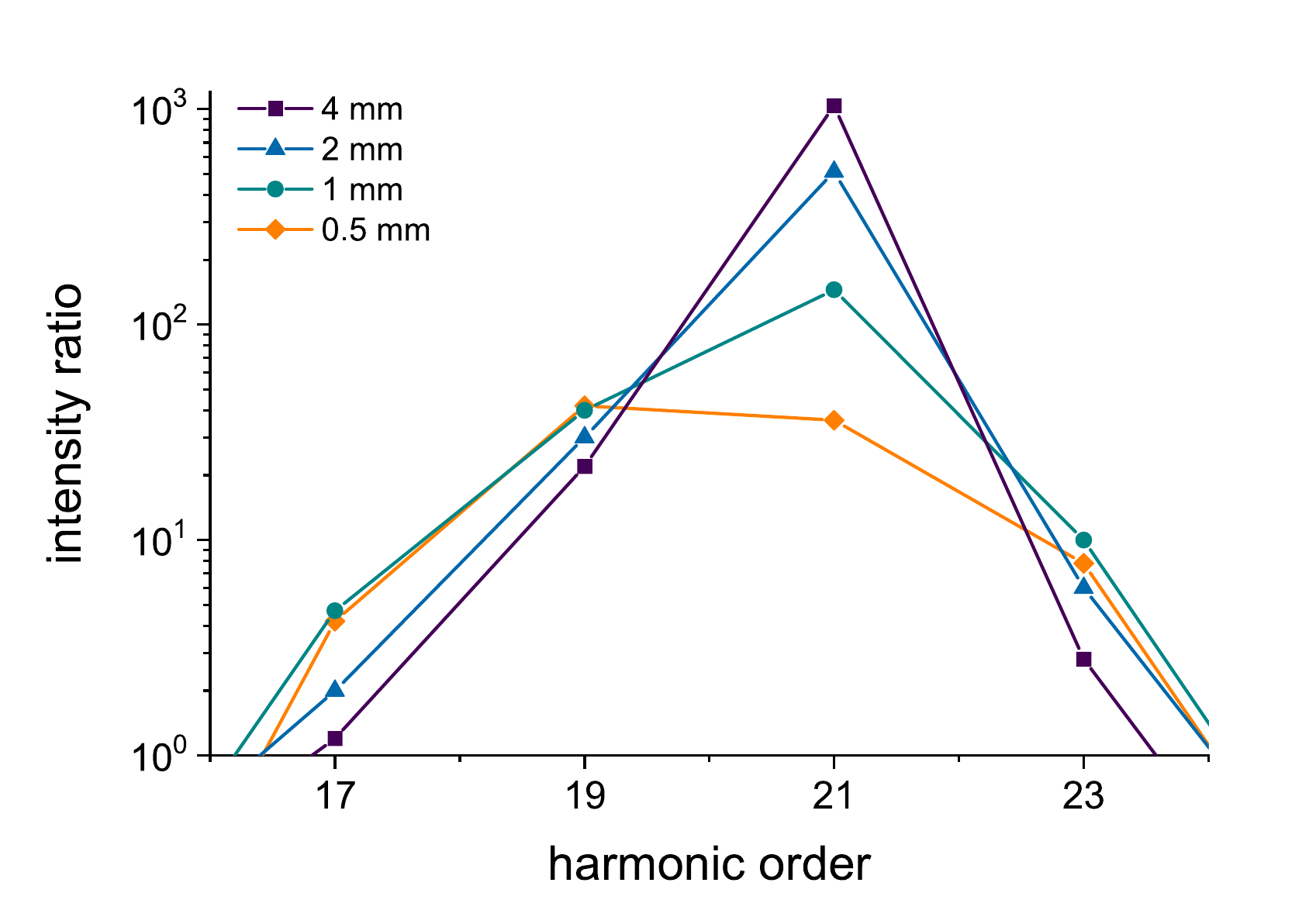}
\caption{Gain or ratio between the peak intensities for the HFM ($m=-2$) and HHG signals as functions of $q$ after propagating in plasma. The intensity of the fundamental field is $\SI{1.2e14}{W/cm^2}$, the wavelength is \SI{800}{nm}, the pulse duration is \SI{50}{fs}. The weak-field frequency is $\omega_1=2/21 \omega_0$ (corresponding to the wavelength $\SI{8.4}{\micro m}$), its intensity is $2\%$ of the fundamental one, the pulse duration is \SI{700}{fs}. The initial ionization degree is unity and the plasma density is $\SI{1.0e18}{cm^{-3}}$.
}
\label{plasma_gain}
\end{figure}

Our numerical results agree with these considerations. In~\fig{plasma_gain} we show the gain achieved for HFM generation in plasma for the same frequencies of the fields as for gas generation, the peak laser intensity $\SI{1.2e14}{W/cm^2}$ providing a photoionization degree of $3.4 \%$. In our conditions $L_\mathrm{coh}^{(21)}= \SI{0.07}{mm}$, $L_\mathrm{bs}^{(21)}= \SI{2}{mm}$ and $L_\mathrm{bs}^{(21,-2)}= \SI{9}{mm}$. We present the gain in intensity, which we calculate using the HHG signal averaged over $2 L_\mathrm{coh}^{(q)}$. The gain in energy is similar because the linewidths of high harmonics, the generation of which is limited by phase matching, do not grow with propagation. The HHG signal oscillates as a function of the propagation length. 

In~\fig{plasma_gain} one can see that for moderate propagation distances HFM components with several $q \approx Q$ are generated effectively, thus making the production of intense attopulses feasible using these HFM components. However, for the longer distances all signals with $q \ne Q$ saturate. Only the component with $q=Q$ continues growing until its blue-shift length is achieved. Thus, the gain for this component can be very high, more than two orders of magnitude in our conditions.

\begin{figure}
\includegraphics[width=0.65\linewidth]{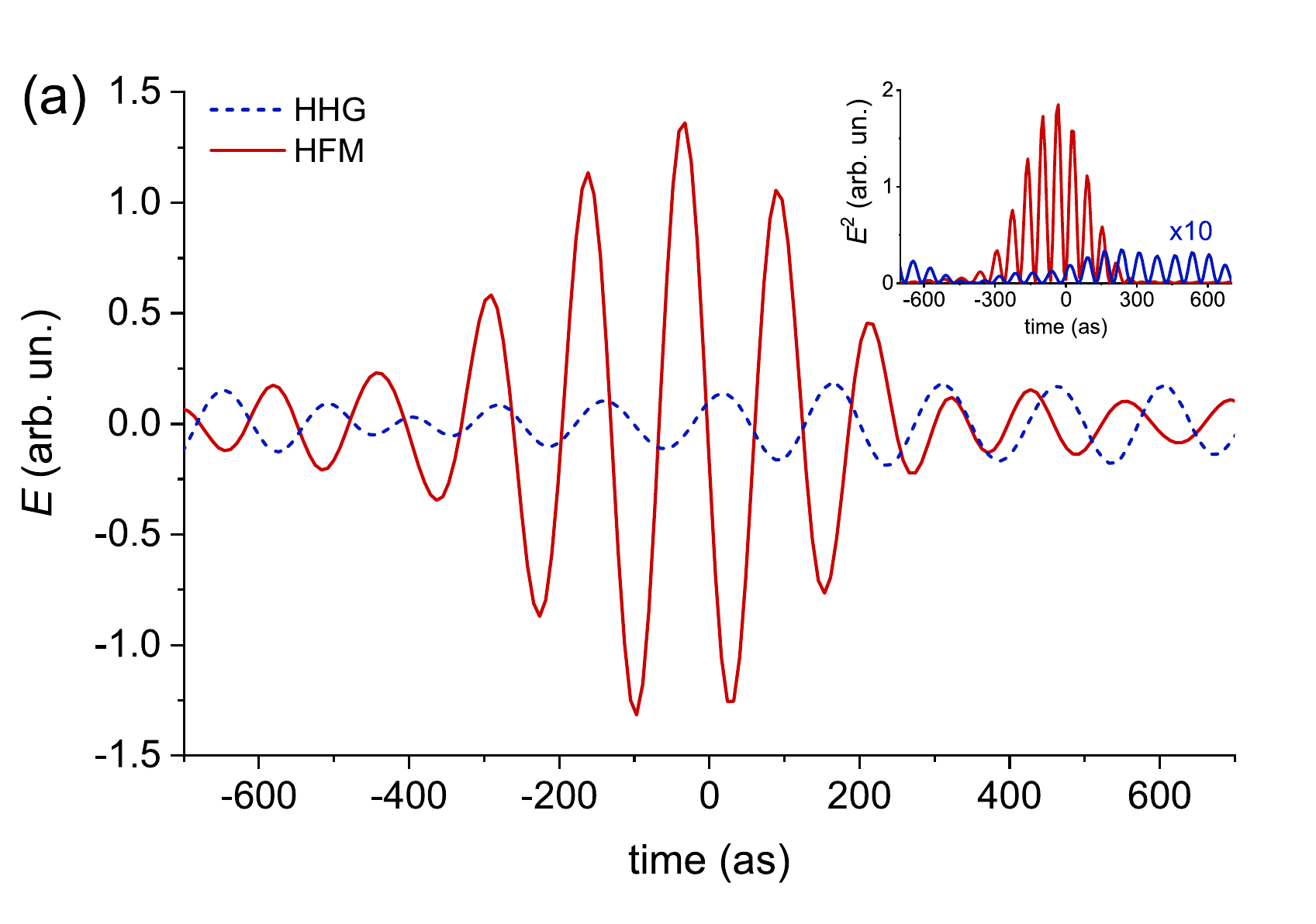}
\includegraphics[width=0.65\linewidth]{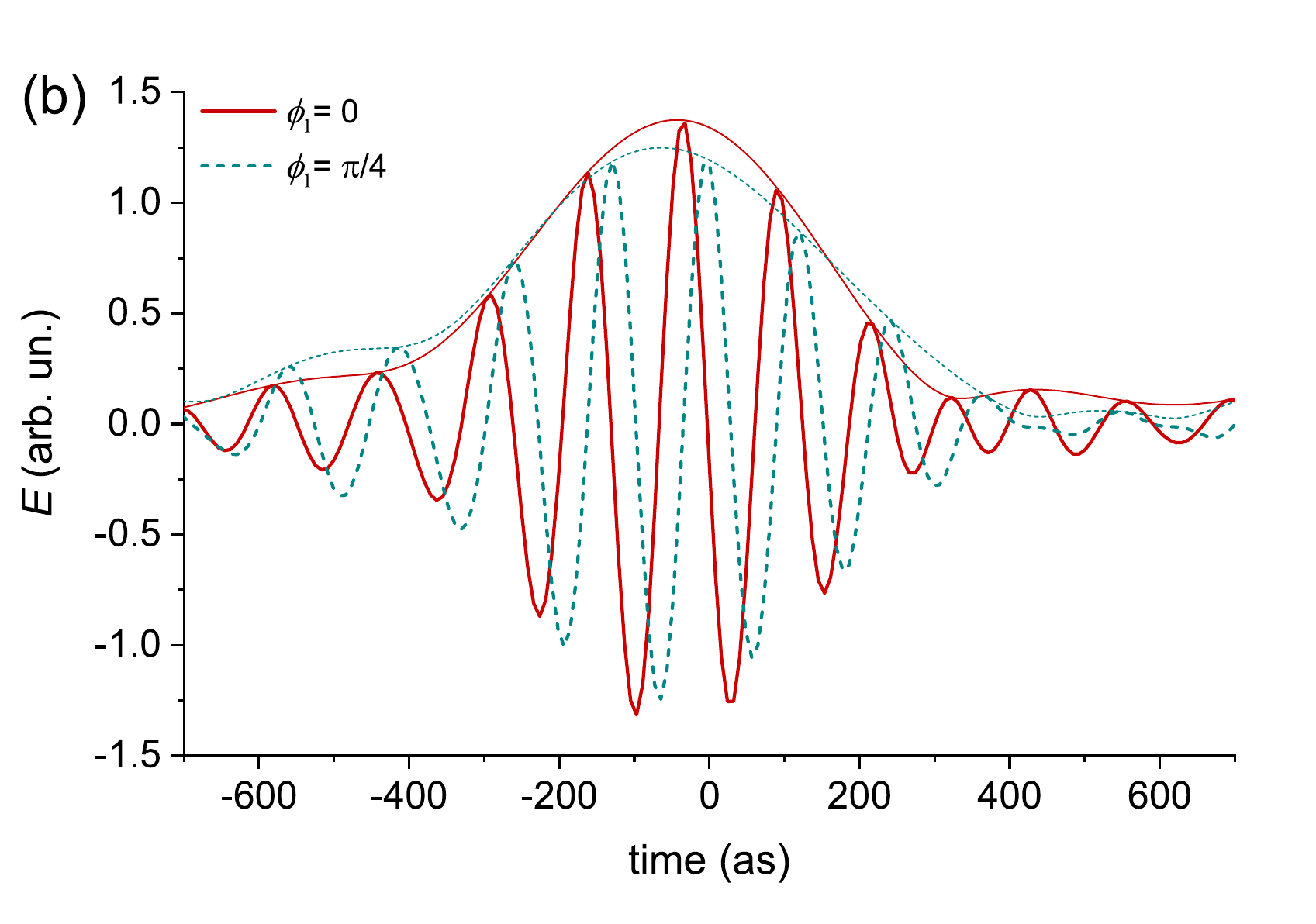}
\caption{(a)~Attosecond pulses obtained via HHG and HFM in plasma, and (b)~the attopulses obtained via HFM for initial phases of the weak generating field differing by $\pi/4$. The propagation distance is \SI{0.5}{mm}.}
\label{plasma_attopulses}
\end{figure}

Attosecond pulses generated via HFM and HHG in plasma for moderate propagation distances are shown in~\fig{plasma_attopulses}. We use XUV higher than H16 to calculate the attosecond pulses. One can see that the HHG field is much weaker and has almost no attosecond modulation. The near absence of the modulation comes from the domination of the single H17~--- the coherence length and thus the harmonic intensity rapidly decrease with the harmonic order, so the lowest harmonic dominates in the chosen spectral region. The full duration at half maximum~(FDHM) of intensity for the attosecond pulse generated via HFM is \SI{310}{as}. It increases for longer propagation distances due to the narrowing of the XUV spectrum with distance discussed above, see~\fig{plasma_gain}. 

Now we look close at the CEP of the attosecond pulse generated via HFM and if it can be controlled by varying the phases of the generating fields. At the microscopic level this control has been studied recently in~\cite{Birulia2022}, where it was shown that the attosecond pulse train formed by the HFM components has a constant carrier-envelope offset (CEO), i.e., a constant shift of the CEP from each attosecond pulse to the next. This CEO naturally corresponds to the shift from the HHG comb to the HFM sidebands, in the same way that it can be observed as a frequency shift in frequency-comb spectroscopy~\cite{Pupeza2021}.

Our current simulations show that this controllability is also the case for the macroscopic response. For example, in~\fig{plasma_attopulses} one can notice that when the initial phase of the weak field changes by $\pi/4$, the phase of the XUV field obtained from the HFM components with $m=-2$ changes by $-2 \times \pi/4 = - \pi/2$. The envelope of the attosecond pulse does not change, so the variation of the carrier phase means the variation of the CEP of the attosecond pulse. We would like to stress that this is a non-trivial result because the phases of the generating fields change significantly during the propagation.

\section{Discussion}\label{Discussion} 
Commonly, for HHG experiments the duration of the driving pulses does not go over a hundred femtoseconds. The application of longer laser pulses leads to ionization of the medium and, as a result, to poor phase matching, even before achieving high laser intensity for driving the generation. This sets a requirement on the laser pulse duration, which limits the choice of laser types. 

The HFM process occurs effectively even under deep ionization of the medium. Our numerical results show that it has reasonable efficiency for long durations of the generating pulses; thus, the choice of the laser sources is much wider. Moreover, one can use generating fields obtained as second or third harmonics of the fundamental pulse, as well as its parametric frequency conversion, without further post-compression of the pulse.

Our numerical calculation of the microscopic response underestimates the XUV absorption (because we use the SAE TDSE). This leads to some overestimation of the HHG efficiency in our calculations. However, this imperfection is less pronounced for the HFM process because it takes place under deeper ionization, thus under lower density of absorbing atoms, since XUV absorption by ions is much weaker. As a result, the ratio between the HFM and HHG efficiency is underestimated in our calculations.

The narrow XUV lines achieved in HFM in gases can provide a precise frequency standard in the XUV domain. Namely, we found in our numerical calculations that the linewidth of the HFM component with $q=17$, $m=-2$ is \SI{12}{meV}, which corresponds to a relative bandwidth of $\delta \omega / \omega = 4.6 \times 10^{-4}$ and which is approximately 6.5 times less than the one achieved via HHG using narrow window resonance in Ref.~\cite{Rothhardt2014}. Moreover, the central frequency of the line is defined very precisely from the initial wavelengths of the generating fields. Thus, HFM can provide an extension of optical frequency standards to the XUV range using only common table-top equipment and without requiring a femtosecond enhancement cavity setup in vacuum~\cite{Pupeza2021}.  

The phase matching of the HFM process can be understood also as quasi-phase matching of HHG: the microscopic response of the $q$-th harmonic is spatially modulated with the additional field. Experimentally quasi-phase-matched HHG has been realized using several gas jets~\cite{Seres2007, Pirri2008, Pullen2013} or several plasma plumes~\cite{Ganeev2014, Strelkov2017}, as well as  micro-fluidic devices~\cite{Ciriolo2020}. However, the size and number of the jets is limited by technical limitations. The absence of such a limitation for the HFM process defines its high efficiency, especially for generation in plasma. Moreover, in plasma the microscopic response at certain XUV frequencies is resonantly enhanced. 

The investigation of an intense monochromatic XUV source based on phase-matched generation of a single HFM component combined with resonant enhancement of the microscopic response for this frequency is a natural outlook of our study.

\section{Conclusion}
In this paper we study the macroscopic properties of HFM, focusing at the practically important case, when one of the fields has much lower frequency and much lower intensity than the strong laser driver. 

We analytically find the length scales limiting the quadratic growth of the macroscopic HFM signal with propagation length. For several HFM components with orders close to $Q$ (defined by the frequencies of the generating fields in Eq.~\eqref{freq1}), these length scales are much longer than the ones for HHG. This provides the higher HFM macroscopic signal despite the lower microscopic response for this process. 

This conclusion is confirmed by our numerical simulations of HFM and HHG propagation based on the coupled solution of the 1D reduced propagation equation and 3D TDSE for the calculation of the microscopic response. 

Moreover, the ratio between the HFM and HHG intensities found in our numerical simulations is between 1 and 4 for generation in a gas and up to three orders of magnitude for generation in plasma. We also find that this ratio is higher for longer generating pulses. 

We also find that HFM for the case of a long fundamental pulse leads to very narrow XUV lines. In the numerical calculations, we find the HFM linewidth $\delta \omega / \omega = 4.6 \times 10^{-4}$. Thus, the HFM process can provide a frequency standard in the XUV range.

Finally, we show that the group of HFM components effectively generated due to macroscopic effects provide a train of attosecond pulses. The CEP of the attosecond pulse can be easily controlled by tuning the phase of one of the generating fields.

\section*{Acknowledgments}
This study was funded by RSF (grant No 22-22-00242). 

\bibliography{lit}

\end{document}